\magnification = 1200
\baselineskip = 15 pt

\def \noi {\noindent}

\def \P {{\bf P}}
\def \E {{\bf E}}

\def \la {\lambda}
\def \a {\alpha}

\def \R {{\bf R}}

\def\sqr#1#2{{\vcenter{\vbox{\hrule height.#2pt\hbox{\vrule width.#2pt height#1pt \kern#1pt\vrule width.#2pt}\hrule height.#2pt}}}}

\def \square{\hfill\mathchoice\sqr56\sqr56\sqr{4.1}5\sqr{3.5}5}

\def \qed {$\square$ \medskip}

\def \sect#1{\bigskip \noindent {\bf  #1} \medskip}
\def \subsect#1{\medskip \noindent{\it #1} \medskip}
\def \th#1#2{\medskip \noindent {\bf  Theorem #1.}   \it #2 \rm \medskip}
\def \prop#1#2{\medskip \noindent {\bf  Proposition #1.}   \it #2 \rm \medskip}
\def \cor#1#2{\medskip \noindent {\bf  Corollary #1.}   \it #2 \rm \medskip}
\def \pf {\noindent  {\it Proof}.\quad }
\def \lem#1#2{\medskip \noindent {\bf  Lemma #1.}   \it #2 \rm \medskip}
\def \ex#1{\medskip \noindent {\bf  Example #1.}}
\def \rem#1 {\medskip \noi {\bf Remark #1.  }}

\centerline{\bf  Minimizing the Probability of Ruin when Consumption is Ratcheted} \bigskip \bigskip

\centerline{Erhan Bayraktar} \medskip

\centerline{Virginia R. Young} \bigskip \bigskip

\centerline{Department of Mathematics, University of Michigan}\medskip
\centerline{Ann Arbor, Michigan, 48109} \bigskip \bigskip

\centerline{Version: 28 May 2008}

\vfill
\eject

\centerline{\bf  Minimizing the Probability of Ruin when Consumption is Ratcheted} \bigskip

\noindent{\bf  Abstract:}  We assume that an agent's rate of consumption is {\it ratcheted}; that is, it forms a non-decreasing process.   Given the rate of consumption, we act as financial advisers and find the optimal investment strategy for the agent who wishes to minimize his probability of ruin.

\medskip


\medskip


\medskip

\noi{\bf Journal of Economic Literature Classification:} G11 (primary); D14, D81 (secondary).

\sect{1. Introduction}

Individual consumers and especially beneficiaries of endowment funds generally employ consumption strategies such that consumption never decreases, or at least, they wish to do this.  In light of Dybvig (1995), we refer to such a consumption pattern as {\it ratcheted}.  In this paper, we take the viewpoint of a financial adviser and find the optimal investment strategy for an  individual or an endowment fund that wishes to minimize the probability of running out of money either before dying or before the organization holding the endowment fails (due to causes other than the ruin of the fund itself), respectively.  We refer to this individual or endowment fund as the {\it agent} and use the pronoun {\it he} to refer to this agent.

Dybvig (1995) maximizes expected discounted utility of consumption under power and logarithmic utility.  He constrains wealth to be such that ruin is impossible; therefore, he assumes that initial wealth $W_0 \ge c_0/r$, in which $c_0$ is the rate of consumption at time 0 and $r$ is the rate of return on a riskless asset.  Thereafter, the agent consumes a constant multiple of maximum wealth.   In other words, Dybvig considers those agents whom one might call ``wealthy'' because their initial wealth is great enough to support their desired consumption without ruining.

By contrast to Dybvig, we consider agents with initial wealth $W_0 < c_0/r$, that is, those who are not rich enough to support their desired consumption.  Note that for such an agent, under non-decreasing consumption, there is a positive probability of ruin; therefore, Dvbvig's problem is infeasible under the constraint that wealth remain positive.  We assume that these poorer agents mimic their wealthier counterparts by consuming an increasing function of maximum wealth.  In particular, we consider the case for which the rate of consumption is a constant proportion of maximum wealth, as in Dybvig's solution for wealthy individuals, and we show that the agent will never allow his wealth to exceed his current maximum.

We take the point of view of financial advisers and advise those agents as to how to invest their wealth to minimize their probability of ruin.  In other words, because we are acting as financial advisers, we take the agent's consumption as given, and in light of that consumption, we advise the agent as to how to invest his wealth.

Young (2004) solves the problem of minimizing the probability of ruin when the rate of consumption is constant or a multiple of wealth.   Bayraktar and Young (2007) consider the same objective function given a rate of consumption that is a piecewise linear function of wealth, and they show that the corresponding investment strategy for the individual is identical to the strategy if that individual were to maximize expected utility of lifetime consumption under HARA utility.  Then, Bayraktar and Young (2008) minimize the probability of lifetime ruin when consumption follows a geometric Brownian motion.  By contrast, in this paper, we consider a rate of consumption that is a function of {\it maximum} wealth; thereby, the rate of consumption is ratcheted.

The rest of the paper is organized as follows: In Section 2.1, we introduce the financial market and define the problem of minimizing the probability of ruin when consumption is ratcheted.  In Section 2.2, we prove a verification theorem for the minimum probability of ruin.  In Sections 3 and 4, we consider various cases for the values of wealth of the agent and the parameters of the model and solve the problem in those cases.

\sect{2. Probability of Ruin}

In Section 2.1, we present the financial market and define the problem of minimizing the probability of ruin.   In Section 2.2, we prove a verification theorem for the minimum probability of ruin.

\subsect{2.1. Financial Market and Probability of Ruin}

In this section, we first present the financial ingredients that make up the agent's wealth, namely, consumption, a riskless asset, and a risky asset. We, then, determine the minimum probability of ruin.  We assume that the agent invests in a riskless asset whose price at time $t$, $X_t$, follows the process $dX_t = rX_t dt, X_0 = x > 0$, for some fixed rate of interest $r > 0$.  Also, the agent invests in a risky asset whose price at time $t$, $S_t$, follows geometric Brownian motion given by

$$\left\{ \eqalign{dS_t &= \mu S_t dt + \sigma S_t dB_t, \cr
S_0 &= S > 0,} \right. \eqno(2.1)$$
\noindent in which $\mu > r$, $\sigma > 0$, and $B$ is a standard Brownian motion with respect to a filtration of a probability space $(\Omega, {\cal F}, \P)$.  Let $W_t$ be the wealth at time $t$ of the agent, and let $\pi_t$ be the amount that the decision maker invests in the risky asset at that time.  It follows that the amount invested in the riskless asset is $W_t - \pi_t$.  Also, define the maximum wealth $M_t$ at time $t$ by
$$
M_t = \max \left[ \sup_{0 \le s \le t} W_s, \; M_0\right], \eqno(2.2)
$$
\noindent in which we include $M_0 = m$ (possibly different from $W_0 = w$) to allow for the agent to have a financial past.

To model ratcheted consumption, we assume that the rate of consumption $c = c(m)$ is a positive, increasing $C^1$ function of maximum wealth.  Thus, wealth follows the process
$$
\left\{ \eqalign{dW_t &= [rW_t + (\mu - r) \pi_t  - c(M_t)] dt + \sigma \pi_t dB_t, \cr
W_0 &= w, \quad M_0 = m.} \right. \eqno(2.3)
$$

By ``ruin,'' we mean that the agent's wealth reaches $0$ before the agent ``dies,'' that is, before the individual dies physically or before the organization holding the endowment fund no longer exists.  Let $\tau_0$ denote the first time that wealth equals $0$, and let $\tau_d$ denote the random time of death of our agent.  We assume that $\tau_d$ is exponentially distributed with parameter $\lambda$ (that is, with expected time of death equal to $1/\lambda$); this parameter is also known as the {\it hazard rate} of the agent.

Denote the minimum probability that the agent outlives his wealth by $\psi(w, m)$, in which the argument $w$ indicates that one conditions on the agent possessing wealth $w$ and maximum wealth $m$ at the current time.  Thus, $\psi$ is the minimum probability that $\tau_0 < \tau_d$, in which one minimizes with respect to admissible investment strategies $\pi$.  A strategy $\pi$ is {\it admissible} if it is ${\cal F}_t$-progressively measurable (in which ${\cal F}_t$ is the augmentation of $\sigma(W_s: 0 \le s \le t)$) and if it satisfies the integrability condition $\int_0^t \pi_s^2 \, ds < \infty$ almost surely for all $t \ge 0$.  Thus, $\psi$ is formally defined by
$$
\psi(w, m) = \inf_{\pi} \P \left[\tau_0 < \tau_d \vert W_0 = w, M_0 = m \right], \eqno(2.4)
$$
for $w \le m$.  Here, $\P^{w, m}$ indicates the probability conditional on $W_0 = w$ and $M_0 = m$.  Below, we similarly write $\E^{w, m}$ for the conditional expectation.

\rem{2.1} Note that we can express $\psi$ as follows.  This alternative representation will prove useful in proving the verification theorem in the next section.
$$
\eqalign{\psi(w, m) &= \inf_\pi \E^{w, m} \left[  \int_0^\infty  {\bf 1}_{\{ \tau_0 \le t \}} \, \la \, e^{-\la t} \, dt \right] \cr
&= \inf_\pi \E^{w, m} \left[  \int_{\tau_0}^\infty  {\bf 1}_{\{ \tau_0 < \infty \}} \, \la \, e^{-\la t} \, dt \right] \cr
&= \inf_\pi \E^{w, m} \left[  e^{-\la \tau_0}  {\bf 1}_{\{ \tau_0 < \infty \}} \right] \cr
&= \inf_\pi \E^{w, m} \left[  e^{-\la \tau_0} \right].} \eqno(2.5)
$$

\subsect{2.2. Verification Theorem}

In this section, we prove a verification theorem for the minimum probability of ruin.  First, note that if $w \ge c(m)/r$, then ruin is impossible.  Indeed, if the agent puts all his wealth $W_t$ in the riskless asset for $t \ge 0$ and consumes the investment earnings $r W_t$, then wealth will steadily increase (or not decrease) until the first time $s$ at which $W_s = c(M_s)/r$.  If $s$ is finite, then $W_t = W_s$ almost surely for $t \ge s$.  In other words, if $w \ge c(m)/r$, then $W_t \ge w$ almost surely for all $t \ge 0$, so wealth never decreases below $c(m)/r$, much less reach $0$.

Next, define the differential operator ${\cal L}^\a$ for $\a \in \R$ by
$$
{\cal L}^\a f = (rw + (\mu - r) \a - c(m)) f_w + {1 \over 2} \sigma^2 \a^2 f_{ww} - \la f, \eqno(2.6)
$$
in which $f = f(w, m)$ is twice-differentiable with respect to its first variable.

\th{2.1} {Let ${\bf D} = \{(w,m) \in {\bf R} \times {\bf R}: w \le m \}$. Suppose $h: {\bf D} \rightarrow
{\bf R}$ is a bounded, continuous function that satisfies the following conditions:
\item{$(i)$} $h(\cdot, m) \in C^2((0, m \wedge c(m)/r))$ is a non-increasing, convex function;
\item{$(ii)$} $h(w, \cdot)$ is continuously differentiable$;$
\item{$(iii)$} $h_m(m, m) \ge 0;$
\item{$(iv)$} $h(w, m) = 1$ if $w \le 0;$
\item{$(v)$} $h(w, m) = 0$ if $w \ge c(m)/r;$
\item{$(vi)$} ${\cal L}^\a h \ge 0$ for all $\a \in {\bf R};$}

\noi{\it Moreover, suppose there is an admissible investment strategy $\beta$ such that on $\bf D,$ $h(w, m) = \P^{w, m} \left[\tau_0 < \tau_d \right]$ when $W = W^\beta$.  Then, $h(w, m) = \psi(w, m)$ on $\bf D$ and $\beta$ is an optimal investment strategy.}

\medskip

\pf Assume that $h$ satisfies the conditions specified in the statement of this theorem.  Let $\pi: {\bf D} \rightarrow {\bf R}$ be a function, and let $W^\pi$ and $M^\pi$ denote the wealth and the maximum wealth, respectively, when the agent uses the investment policy $\pi_t = \pi(W_t, M_t)$.  Assume that this investment policy is admissible.  

Define $\tau_n = \inf \{t \ge 0:  \int_0^t \pi^2_s \, ds \ge n \}$ and $\tau = \tau_0 \wedge \tau_n$.  By applying It\^{o}'s formula to $e^{-\la t} h(w, m)$, we have
$$
\eqalign{e^{-\la \tau} h(W^\pi_{\tau}, M^\pi_{\tau}) &= h(w, m) + \int_0^{\tau} e^{-\la t} \, h_w(W^\pi_t, M^\pi_t) \, \sigma \, \pi_t \, dB_t  \cr
& \quad +\int_0^{\tau} e^{-\la t} \, {\cal L}^\pi h(W_t, M_t) \, dt  + \int_0^{\tau} e^{-\la t} \, h_m (W^\pi_t, M^\pi_t) \, dM^\pi_t.} \eqno(2.7)
$$

It follows from the definition of $\tau_n$ that
$$
\E^{w, m} \left[\int_0^{\tau} e^{-\la t} \, h_w(W^\pi_t, M^\pi_t) \, \sigma \, \pi_t \, dB_t \right] = 0. \eqno(2.8)
$$
Also, the second integral in (2.7) is non-negative because of condition (vi) of the theorem.  Finally, the third integral  is non-negative almost surely because $dM_t$ is non-zero only when $M_t = W_t$ and $h_m(m,m) \ge 0$.   Here, we also used the fact that $M$ is non-decreasing, therefore the first variation process associated with it is finite almost surely, to conclude that the cross variation of $M$ and $W$ is zero almost surely.  Thus, we have
$$
\E^{w,m} [e^{-\la \tau} h(W^\pi_{\tau}, M^\pi_{\tau})] \ge h(w, m).  \eqno(2.9)
$$

Because $h$ is bounded by assumption, it follows from the dominated convergence theorem that
$$
\E^{w,m}[e^{-\la \tau_0} h(W^\pi_{\tau_0}, M^\pi_{\tau_0})] \ge h(w,m).  \eqno(2.10)
$$
Since $W^\pi_{\tau_0} = 0$, it follows from condition (iv) of the theorem that
$$
\E^{w,m}[e^{-\la \tau_0} ] \ge h(w,m).  \eqno(2.11)
$$
By taking the infimum over admissible investment strategies, and by applying the representation of $\psi$ from (2.5), we obtain $\psi \ge h$ on $\bf D$.

From the definition of $\psi$ in (2.4) and from the fact that $h$ is the probability of ruin corresponding to an admissible investment strategy, we know that $\psi \le h$.  Thus, $h = \psi$ on $\bf D$.  \qed

In the next two sections, we use this verification theorem to solve for the minimum probability of ruin $\psi$.

\sect{3.  Solving for $\psi$ when $w < c(m)/r \le m$}

From the discussion at the beginning of Section 2.2, recall that if $w \ge c(m)/r$, then $\psi$ is identically $0$.  Therefore, we only need to consider $0 < w < c(m)/r$.  In this section, we determine $\psi$ when $0 < w < c(m)/r \le m$.  In the next section, we determine $\psi$ when $0 < w \le m < c(m)/r$.

Define an investment strategy $\pi$ as a feedback control as follows:
$$
\pi_t = {\mu - r \over \sigma^2} \cdot {1 \over \gamma - 1} \, \left( {c(M^\pi_t) \over r} -  W^\pi_t \right), \eqno(3.1)
$$
in which $\gamma$ is defined by
$$
\gamma = {1 \over 2r} \left[ (r + \lambda + \delta) + \sqrt{(r + \lambda + \delta)^2 - 4r \lambda} \right] > 1, \eqno(3.2)
$$
with
$$
\delta = {1 \over 2} \left( {\mu - r \over \sigma} \right)^2.  \eqno(3.3)
$$
Recall that $W^\pi$ and $M^\pi$ denote the wealth and maximum wealth, respectively, under the investment strategy $\pi$. 

One can show that $W^\pi$ follows the process
$$
dW^\pi_t = \left( {c(M^\pi_t) \over r} - W^\pi_t \right) \left\{  \left( -r + {2 \delta \over \gamma - 1} \right) dt + {\mu - r \over \sigma} \cdot {1 \over \gamma - 1} dB_t \right\}.  \eqno(3.4)
$$
Note that because $W_0 = w$ and $M_0 = m$ satisfies $w < c(m)/r \le m$, under this investment strategy, $W^\pi_t < c(m)/r$ almost surely for all $t \ge 0$.  Thus, $M^\pi_t = m$ almost surely for all $t \ge 0$.  In other words, the agent's maximum wealth is constant and the corresponding consumption rate equals the constant $c(m)$.  From Young (2004), we know that the probability of ruin in this case is given by
$$
\phi(w, m) = \left( 1 - {r w \over c(m)} \right)^\gamma, \quad 0 < w < c(m)/r.  \eqno(3.5)
$$
We write $\phi$ for this probability of ruin because we do not yet know that it is the minimum.  If $w \le 0$, then we define $\phi$ to be identically $1$.  In the next theorem, we show that in this case, $\phi$ is the minimum probability of ruin.

\th{3.1} {When $w < c(m)/r \le m,$ the minimum probability of ruin $\psi$ equals $\phi,$ which is given by $(3.5)$ when $w > 0$ and equals $1$ when $w \le 0$.  The corresponding optimal investment strategy for $w > 0$ is given in feedback form by $(3.1)$.}

\pf  To prove this statement, we show that $\phi$ satisfies the conditions of Theorem 2.1.  It is clear that $\phi$ satisfies conditions (i), (ii), and (iv) of Theorem 2.1.  We do not need to consider condition (v) because under the investment strategy defined in (3.1), it is impossible for wealth to reach $c(m)/r$ when initial wealth $W_0 = w < c(m)/r$.  Also, condition (vi) is satisfied because $\phi$ solves the Hamilton-Jacobi-Bellman (HJB) equation $\min_\a {\cal L}^\a \phi = 0$.

Finally, consider condition (iv).  For $w > 0$, we compute
$$
\phi_m(m, m) = \gamma \left( 1 - {r w \over c(m)} \right)^{\gamma - 1} {r \, w \, c'(m) \over c^2(m)} > 0, \eqno(3.6)
$$
in which the inequality follows from $c'(m) > 0$. Thus, condition (iv) holds, and we have proved that $\phi \le \psi$.  Because $\phi$ is the probability of ruin for an admissible investment strategy, it must be that $\phi = \psi$.  \qed

Theorem 3.1 tells us that when wealth is less than the ``safe level'' $c(m)/r$, and when that safe level is less than the maximum wealth $m$, in order to minimize the probability of ruin, the agent's wealth cannot reach the safe level and, thereby, cannot reach a new maximum.  The agent effectively treats his consumption rate as constant, and the results of Young (2004) apply.

It follows from (3.1) that as wealth increases towards $c(m)/r$, the amount invested in the risky asset decreases to zero. This makes sense because as the agent becomes wealthier, he does not need to take on as much risk to achieve his fixed consumption rate of $c(m)$.

As an application of Theorem 3.1, we have the following example.

\ex{3.1}  Suppose $c(m) = \rho m$ in which $0 < \rho \le r$, and suppose $0 < w < c(m)/r$.  Thus, we have $0 < w < c(m)/r = \rho m/r \le m$.  Theorem 3.1 implies  that the minimum probability of ruin is given in (3.5) with $c(m)$ replaced by $\rho m$.  We note that $c(m) = \rho m$ is the optimal form of consumption that Dybvig (1995) finds when maximizing discounted utility of consumption under power or logarithmic utility.

\sect{4. Solving for $\psi$ when $m < c(m)/r$}

In the previous section, we showed that it is optimal for $M_t = m$ almost surely for all $t \ge 0$ when $0 < w < c(m)/r \le m$.  In this section, we show that allowing $M$ to increase above $m$ might be optimal when $0 < w \le m < c(m)/r$.  In Section 4.1, we consider a related optimal controller-stopper problem and determine when its Legendre transform is the minimum probability of ruin.  In Section 4.2, we prove properties of $\psi$ and the corresponding optimal investment strategy $\pi^*$ when ratcheting is not optimal.  Then, in Section 4.3, we examine the case for which it is is optimal to allow $M$ to increase above $m$.

\subsect{4.1  A Related Optimal Controller-Stopper Problem}

Define the controlled process $Y^R$ by
$$
Y^R_t = -(r - \la) \, Y^R_t \, dt + {\mu - r \over \sigma} \, Y^R_t \, d \hat B_t + d R_t, \quad Y^R_0 = y > 0, \eqno(4.1)
$$
in which $\hat B$ is a standard Brownian motion with respect to a filtration of a probability space $(\hat \Omega, \hat {\cal F}, \hat \P)$.  Here, $R$ is a right-continuous, non-negative, non-decreasing control that incurs a proportional cost of $m$ when the controller implements it. 

For $y > 0$, define the function $\hat \psi$ by
$$
\hat \psi(y) = \inf_\tau \sup_R \hat \E^y \left[ \int_0^\tau e^{-\la t} \left( c(m) \, Y^R_t \, dt - m \, dR_t \right) + e^{-\la \tau} \right].  \eqno(4.2)
$$
$\hat \psi$ is the value function for an optimal controller-stopper problem.  Indeed, the controller wishes to maximize the discounted (net) running ``penalty'' to the stopper given by $c(m) \, Y^R_t$ in (4.2), net of the controller's proportional cost $m$.  On the other hand, the stopper wishes to minimize the penalty but has to incur the terminal cost of $1$, discounted by $e^{-\la \tau}$, if she stops the game.

Via standard techniques (\O ksendal and Sulem, 2004, Chapter 5), one can show that there exists $y_m > 0$ such that the controller implements the control $R$ in order to keep $y \ge y_m$.  Specifically, if $Y^R_0 = y < y_m$, then the controller immediately moves $Y^R$ to $y_m$ and incurs the cost $m (y_m - y)$.  Thus, for $y < y_m$, we have $\hat \psi(y) = -m(y_m - y) + \hat \psi(y_m)$.  After that, the controller exercises instantaneous control to keep $y \ge y_m$.

Additionally, one can show that there exists $y_0 > y_m$ such that the stopper stops the game immediately if  $Y^R_0 = y \ge y_0$, and if $y < y_0$, then she stops when $Y^R$ reaches $y_0$.  Thus, if $y \ge y_0$, we have $\hat \psi(y) = 1$.

Moreover, $\hat \psi$ is concave on $\R^+$ and is the unique classical solution of the following free-boundary problem on $(y_m, y_0)$:
$$
\left\{ \eqalign{&\delta y^2 f'' - (r - \la) y f' - \la f + c(m) y = 0, \quad y_m < y < y_0; \cr
&f(y_0) = 1, \quad f'(y_0) = 0; \cr
&f'(y_m) = m, \quad f''(y_m) = 0.} \right.  \eqno(4.3)
$$
Because $\hat \psi$ is concave, we can define its convex dual by the Legendre transform.  First, we solve (4.3) for $\hat \psi$, then we show how to compute its convex dual, and finally we determine when that convex dual is the minimum probability of ruin in the case for which $m < c(m)/r$.

The general solution of the ODE in (4.3) is given by
$$
\hat \psi(y) = D_1 \, y^{B_1} + D_2 \, y^{B_2} + {c(m) \over r} \, y, \eqno(4.4)
$$
in which
$$
B_1 = {1 \over 2 \delta} \left[ (r - \lambda + \delta) + \sqrt{(r - \lambda + \delta)^2 + 4 \lambda \delta} \right] > 1, \eqno(4.5)
$$
and
$$
B_2 = {1 \over 2 \delta} \left[ (r - \lambda + \delta) - \sqrt{(r - \lambda + \delta)^2 + 4 \lambda \delta} \right] < 0. \eqno(4.6)
$$
The constants $D_1$ and $D_2$ are to be determined.  Note that $B_1 = \gamma/(\gamma - 1)$.

The four boundary conditions imply that
$$
D_1 \, y_0^{B_1} + D_2 \, y_0^{B_2} + {c(m) \over r} \, y_0 = 1,  \eqno(4.7)
$$
$$
D_1 \, B_1 \, y_0^{B_1 - 1} + D_2 \, B_2 \, y_0^{B_2 - 1} + {c(m) \over r} = 0,  \eqno(4.8)
$$
$$
D_1 \, B_1 \, y_m^{B_1 - 1} + D_2 \, B_2 \, y_m^{B_2 - 1} + {c(m) \over r} = m,  \eqno(4.9)
$$
and
$$
D_1 \, B_1 (B_1 - 1) \, y_m^{B_1 - 2} + D_2 \, B_2 (B_2 - 1) \, y_m^{B_2 - 2}  = 0,  \eqno(4.10)
$$
which gives us four equations to determine the four unknowns $D_1$, $D_2$, $y_m$, and $y_0$.  Solve (4.9) and (4.10) for $D_1$ and $D_2$ to obtain
$$
D_1 = - \, {1 - B_2 \over B_1(B_1 - B_2)} \left( {c(m) \over r} - m \right) {1 \over y_m^{B_1 - 1}} < 0,  \eqno(4.11)
$$
and
$$
D_2 = - \, {B_1 - 1 \over B_2(B_1 - B_2)} \left( {c(m) \over r} - m \right) {1 \over y_m^{B_2 - 1}} > 0.  \eqno(4.12)
$$

Next, substitute these expressions for $D_1$ and $D_2$ into (4.8) to get the following equation for $y_0/y_m$:
$$
{1 - B_2 \over B_1 - B_2} \left( {y_0 \over y_m} \right)^{B_1 - 1} + {B_1 - 1 \over B_1 - B_2} \left( {y_0 \over y_m} \right)^{B_2 - 1} = {c(m) \over c(m) - rm}.  \eqno(4.13)
$$
Equation (4.13) has a unique solution $y_0/y_m > 1$ because (a) when $y_0/y_m = 1$, the left-hand side of (4.13) equals $1$, which is less than $c(m)/(c(m) - rm)$; (b) as $y_0/y_m \to \infty$, the left-hand side approaches $\infty$; and (c) the left-hand side is strictly increasing with respect to $y_0/y_m$.

Once we solve (4.13) for $y_0/y_m$, we can get $y_0$ from (4.7) after substituting for $D_1$ and $D_2$ in terms of $y_m$.  Indeed,
$$
{1 \over y_0} = {c(m) \over r} - \left( {c(m) \over r} - m \right) \left[ {1 - B_2 \over B_1 (B_1 - B_2)} \left( {y_0 \over y_m} \right)^{B_1 - 1} + {B_1 - 1 \over B_2 (B_1 - B_2)} \left( {y_0 \over y_m} \right)^{B_2 - 1} \right].  \eqno(4.14)
$$
Finally, we get $y_m$ by computing $y_m = y_0/(y_0/y_m)$, and from $y_m$, we get $D_1$ and $D_2$ from (4.11) and (4.12), respectively.

Because $\hat \psi$ is concave, we can define its convex dual $\Psi$ by
$$
\Psi(w) = \max_y \left( \hat \psi(y) - wy \right). \eqno(4.15)
$$
Note that $\Psi$ implicitly depends on $m$ because $\hat \psi$ depends on $m$ in (4.2).  In the next lemma, we show that $\Psi$ is a probability of ruin under a restriction on the admissible investment strategies.

\lem{4.1} {If $m < c(m)/r,$ then $\Psi$ is the minimum probability of ruin under the restriction that $M_t = m$ almost surely for $t \ge 0,$ that is, wealth may not grow larger than $m$.}

\pf  From (4.15), it follows that the critical value $y^*$ solves $w = \hat \psi'(y)$; thus, given $w$, we have $y^* = I(w)$, in which $I$ is the inverse function of $\hat \psi'$.  Therefore, $\Psi(w) = \hat \psi(I(w)) - w I(w)$.  By differentiating this expression for $\Psi$, we obtain $\Psi'(w) = \hat \psi'(I(w)) I'(w) - I(w) - w I'(w) = - I(w)$; thus, $y^* = - \Psi'(w)$.  Also, note that $\Psi''(w) = -1/\hat \psi''(I(w))$.

By substituting $y^* = - \Psi'(w)$ into the free-boundary problem for $\hat \psi$, namely (4.3), it follows that $\Psi$ uniquely solves the following boundary-value problem:
$$
\left\{ \eqalign{&\min_\a {\cal L}^\a f(w) = 0; \cr
&f(0) = 1, \quad \lim_{w \to m-} {f'(w) \over f''(w)} = 0.} \right.  \eqno(4.16)
$$
Note that the condition $\lim_{w \to m-} f'(w)/f''(w) = 0$ is equivalent to $M_t = m$ almost surely for all $t \ge 0$.  Indeed, the optimal investment in the risky asset is given by
$$
\pi^*_t = - \, {\mu - r \over \sigma^2} \, {\Psi'(W^*_t) \over \Psi''(W^*_t)}, \eqno(4.17)
$$
in which $W^*$ is the optimally controlled wealth.  Because $\pi^*_t = 0$ almost surely when wealth reaches $m$ and because the consumption rate $c(m)$ is greater than $rm$, wealth can never become larger than $m$.

Thus, by a verification theorem  similar to Theorem 2.1, we deduce that $\Psi$ is the minimum probability of ruin under the restriction that wealth cannot grow larger than $m$.  \qed

We have the following theorem concerning $\Psi$'s relationship with the (unrestricted) minimum probability of ruin $\psi$.

\th{4.2} {If $m < c(m)/r$ and if $c(m) - m \, c'(m) \le \la/y_0,$ then the minimum probability of ruin $\psi(w, m) = \Psi(w)$ for $0 \le w \le m < c(m)/r$.}

\pf  To prove this theorem, it is sufficient to show that $\Psi$ satisfies the conditions of Theorem 2.1.  It is clear that $\Psi$ satisfies conditions (i) and (ii) of Theorem 2.1.  Also, from (4.15), we compute that $\Psi(0) = 1$, so that condition (iv) holds on the interval under consideration.  Condition (v) is irrelevant because we will show that wealth never increases beyond $m$, and we are assuming that $m < c(m)/r$.  Condition (vi) is satisfied because $\Psi$ solves the HJB equation $\min_\a {\cal L}^\a \Psi = 0$.

It remains to show that $\Psi$ satisfies condition (iv).  Now, $\Psi_m(m, m) \ge 0$ if and only if $\hat \psi_m(y_m, m) \ge 0$, in which we make explicit the dependence of $\Psi$ and $\hat \psi$ on $m$.  It is straightforward to show that
$$
\hat \psi_m(y_m, m) = {c(m) - rm \over \la} \; {\partial y_m \over \partial m} + {c'(m) - r \over \la} \; y_m + y_m.  \eqno(4.18)
$$
Through a truly tedious calculation, one can show that
$$
{\partial y_m \over \partial m} = - y_m \left( y_0 \, {c(m) - m \, c'(m) \over c(m) - rm} + {c'(m) - r \over c(m) - rm} \right), \eqno(4.19)
$$
from which it follows that $\Psi_m(m, m) \ge 0$ if and only if $c(m) - m \, c'(m) \le \la/y_0$.  Thus, because that $\Psi$ equals a probability of ruin function as shown in Lemma 4.1, it follows from Theorem 2.1 that $\Psi = \psi$.  \qed

We have a corollary that follows immediately from Theorem 4.2 and the work at the beginning of this section.

\cor{4.3} {If $m < c(m)/r$ and if $c(m) - m \, c'(m) \le \la/y_0,$ then for $0 < w < m,$ the minimum probability of ruin $\psi$ is given by
$$
\psi(w, m) = D_1 \, y^{B_1} + D_2 \, y^{B_2} + \left( {c(m) \over r} - w \right) \, y,  \eqno(4.20)
$$
in which $y$ solves
$$
w = D_1 \, B_1 \, y^{B_1 - 1} + D_2 \, B_2 \, y^{B_2 - 1} + {c(m) \over r}.  \eqno(4.21)
$$
For wealth lying in $(0, m),$ the corresponding optimal investment strategy $\pi^*$ is given in feedback form by $\pi^*_t = \pi^*(W^*_t),$ in which we abuse notation slightly, and in which
$$
\pi^*(w) = - {\mu - r \over \sigma^2} \left( D_1 \, B_1 (B_1 - 1) \, y^{B_1 - 1} + D_2 \, B_2 (B_2 - 1) \, y^{B_2 - 1} \right).  \eqno(4.22)
$$}

\rem{4.1} Theorem 4.2 tells us that if the rate of increase in the consumption function is large enough, specifically larger than $(c(m) -  \la/y_0)/m$, then to minimize the probability of ruin, the agent will not allow his wealth to exceed the current maximum $m$.  From the proof of Lemma 4.1, we learn that $y_0 = -\Psi'(0)$; thus, we can rewrite the condition in the hypothesis of Theorem 4.2 as
$$
c'(m) \ge  {c(m) + \la/\Psi'(0) \over m}.  \eqno(4.23)
$$





\medskip

As an application of Theorem 4.2 and Remark 4.1, we have the following examples and accompanying remarks.

\ex{4.1} Suppose $c(m) = \rho m$ with $\rho > r$.  Then, $m < c(m)/r = \rho m/r$, and $c'(m) = \rho = c(m)/m \ge (c(m) + \la/\Psi'(0))/m$, in which the second inequality follows because $\Psi'(0) \le 0$.  Theorem 4.2 implies that in the optimal investment strategy, the agent will not allow his wealth to exceed the current maximum $m$. 

\rem{4.2} Therefore, from Examples 3.1 and 4.1, we learn that if the rate of consumption equals $\rho m$, then wealth will never exceed the current maximum for an agent who wishes to minimum his probability of ruin.  In the case for which $\rho \le r$, this result follows from the existence of the safe level $c(m)/r = \rho m/r$, which is less than $m$.  In the case for which $\rho > r$, this result follows from the fact that the increase in the rate of consumption is too large.  In other words, despite the agent wanting larger wealth in order to keep wealth away from the ruin level $0$, the ``cost'' of getting larger wealth is a permanent increase in his consumption rate.  This increase is too large if it satisfies (4.23); therefore, he will not allow wealth to increase beyond the current maximum.

We remark again that $c(m) = \rho m$ is the optimal form of consumption that Dybvig (1995) finds when maximizing discounted utility of consumption under power or logarithmic utility.  Thus, if a less-than-wealthy agent mimics his wealthier utility-maximizing counterparts by consuming a constant multiple of maximum wealth, then in order to minimize the probability of ruin, the poorer agent will not allow his wealth to exceed the current maximum.

\ex{4.2} Suppose the rate of consumption $c(m)$ is strictly concave with $c(0) = 0$.  If $c'(0) \le r$, we have $c'(m) \le r$ and $c(m)/r \le m$ for all $m \ge 0$.  Thus, from Theorem 3.1, it follows that the agent will not increase his wealth above the safe level $c(m)/r \le m$ if $w < c(m)/r$.  If $w \ge c(m)/r$, recall from the discussion preceding Theorem 2.1 that ruin is impossible, so we need not consider that case in this example and the next.

On the other hand, if $c'(0) > r$, then there exists $\hat m > 0$ such that $c(\hat m) = r \hat m$, and we have $c(m)/r \le m$ for all $m \ge \hat m$.  It follows that the agent will not increase his wealth above the current maximum $m$ if $m \ge \hat m$ and if $w < c(m)/r$.   If $m < \hat m$, then the agent will allow his wealth to exceed $m$ if $c'(m) < (c(m) + \la/\Psi'(0))/m$, but in this case, wealth will not grow beyond $\hat m$ at the most.

\ex{4.3}  Suppose the rate of consumption $c(m)$ is strictly convex with $c(0) = 0$.  Then, $c'(m) \ge c(m)/m$ for all $m \ge 0$, so that the inequality in (4.23) holds for all $m \ge 0$.  Therefore, if $c(m)/r \le m$, from Theorem 3.1, then the agent will not increase his wealth above the safe level $c(m)/r \le m$ if $w < c(m)/r$ for $m \le \hat m$.  Otherwise, if $m < c(m)/r$, then Theorem 4.2 implies that the agent will not increase his wealth above $m$.

\rem{4.3}  What Theorem 4.2 shows us is that if the agent wishes to ratchet his consumption and allow his maximum wealth to increase {\it without} increasing his probability of ruin, then he simply needs to increase his rate of consumption at a rate less than $(c(m) + \la/\Psi'(0))/m$.  In other words, in addition to giving advice about the optimal investment strategy to minimize the probability of ruin, we can also advise the agent about how he can ratchet his consumption without increasing his probability of ruin.  Note that in this case, the agent's probability of ruin will not equal $\Psi$ because condition (4.23) no longer holds; that is the topic of Section 4.3.

\subsect{4.2  Properties of $\psi$ and $\pi^*$ when Increasing Maximum Wealth is Not Optimal}

In this section, we study properties of the minimum probability of ruin $\psi$ and the corresponding optimal investment strategy $\pi^*$ when the conditions of Theorem 4.2 hold.  Specifically, throughout this section, we assume that $m < c(m)/r$ and if $'(m) \ge  (c(m) + \la/\Psi'(0))/m$, in which $\Psi$ is defined by (4.15).

Recall from Young (2004), that when consumption is constant ($c(m)$ in this case), the minimum probability of ruin $\phi$ is given by
$$
\phi(w) = \cases{ 1, &if $w \ge 0$, \cr
\left( 1 - {r w \over c(m)} \right)^\gamma, &if $0 < w < c(m)/r$, \cr
0, &if $w \ge c(m)/r$,}  \eqno(4.24)
$$
and the corresponding optimal investment strategy when $0 < w < c(m)/r$ is given by the expression in (3.1) with $c(M^\pi_t)$ replaced by $c(m)$ because $M_t = m$ almost surely for all $t \ge 0$.  Now, when $m < c(m)/r$, this investment strategy allows the maximum wealth to increase beyond $m$; thus, this investment strategy is not the optimal one corresponding to the problem considered in Section 4.1.

This observation leads us to ask how the minimum probability of ruin $\psi$ given in Corollary 4.3 compares with $\phi$ in (4.24) and how the investment strategy given in (4.22) compares with the one in (3.1) for $0 < w < m$.

\prop{4.4} {For $0 < w \le m,$ the minimum probability of ruin given in  Corollary $4.3$ is greater than $\left( 1 - {r w \over c(m)} \right)^\gamma,$ in which $\gamma$ is given in $(3.2)$.}

\pf  To prove this result, we show something stronger, namely
$$
\psi_w(w, m) > -  {\gamma \, r \over c(m)} \left( 1 - {rw \over c(m)} \right)^{\gamma - 1}.  \eqno(4.25) 
$$
Because $\psi(0, m) = 1$, the inequality in (4.25) implies that $\psi$ is greater than $\left( 1 - {r w \over c(m)} \right)^\gamma$.

Now, (4.25) is true for all $w \in [0, m)$ if and only if
$$
-y > -  {\gamma \, r \over c(m)} \left( 1 - {r \hat \psi'(y) \over c(m)} \right)^{\gamma - 1},  \eqno(4.26)
$$
is true for all $y \in (y_m, y_0]$.  After substituting for $\hat \psi$ and for $\gamma = B_1/(B_1 - 1)$ and after simplifying, inequality (4.26) is equivalent to
$$
\left( {B_1 - 1 \over B_1} \, {c(m) \over r} \, y_m \right)^{B_1 - 1} < {c(m) - rm \over c(m)} \left[ {1 - B_2 \over B_1 - B_2} + {B_1 - 1 \over B_1 - B_2} \left( {y \over y_m} \right)^{B_2 - B_1} \right].  \eqno(4.27)
$$
Because the right-hand side of (4.27) decreases with respect to $y$, the inequality holds for all $y \in (y_m, y_0)$ if and only if it holds at $y = y_0$.  Recall that $y_0/y_m > 1$ is the unique solution of (4.13).  By substituting that expression into (4.27) with $y = y_0$, we obtain that (4.25) holds for all $w \in [0, m)$ if and only if
$$
\left( {B_1 - 1 \over B_1} \, {c(m) \over r} \, y_0 \right)^{B_1 - 1} < 1,  \eqno(4.28)
$$
or equivalently,
$$
{1 \over y_0} > {B_1 - 1 \over B_1} \, {c(m) \over r}.  \eqno(4.29)
$$
By substituting for $1/y_0$ from (4.14), simplifying, and again using (4.13), one can show that (4.29) is equivalent to
$$
{B_1 - 1 \over B_2} \left( {y_0 \over y_m} \right)^{B_2 - 1} < 0, \eqno(4.30)
$$
which is true because $B_1 > 1$ and $B_2 < 0$.  \qed

This result is consistent with our intuition because both expressions correspond to a minimum probability of ruin with constant consumption $c(m)$.  However, in the ratcheting problem, if wealth increases above $m$, then the consumption increases too much, so there is a constraint on the investment strategy to keep the maximum wealth no greater than $m$.  This constraint does not exist in the minimization problem corresponding to (4.24).

From the proof of Proposition 4.4, we have an immediate corollary.

\cor{4.5} {For $0 \le w < m,$ the difference $\psi(w, m) - \left( 1 - {r w \over c(m)} \right)^\gamma$ is increasing with respect to $w$.}

Thus, not only is $\psi$ greater than the expression in (4.24), the difference increases as wealth increases.  After the results below, we will discuss an explanation of this phenomenon.

\prop{4.6} {For $0 < w < m,$ the optimal investment strategy $\pi^*(w) < {\mu - r \over \sigma^2} \cdot {1 \over \gamma - 1} \, \left( {c(m) \over r} -  w \right)$.}

\pf  Because $\pi^*(w) = -{\mu - r \over \sigma^2} \, {\psi'(w) \over \psi''(w)}$, the inequality is true for all $w \in (0, m)$ if and only if
$$
-y \, \hat \psi''(y) < {1 \over \gamma - 1} \, \left( {c(m) \over r} -  \hat \psi'(y) \right),  \eqno(4.31)
$$
is true for all $y \in (y_m, y_0)$.  After substituting for $\hat \psi$ and simplifying, we observe that the inequality in (4.31) is equivalent to
$$
-(1 - B_2) \left( {y \over y_m} \right)^{B_2 - 1} < (B_1 - 1) \left( {y \over y_m} \right)^{B_2 - 1},  \eqno(4.32)
$$
which is true because the left-hand side is negative and the right is positive.   \qed

Because the difference ${\mu - r \over \sigma^2} \cdot {1 \over \gamma - 1} \, \left( {c(m) \over r} -  w \right) - \pi^*(w)$ is positive on $(0, m)$, and because it is related to the difference in the corresponding probabilities of ruin, we ask if the difference in the investment strategies is monotone with respect to $w$.  The next proposition proves that this is the case.

\prop{4.7} {The difference ${\mu - r \over \sigma^2} \cdot {1 \over \gamma - 1} \, \left( {c(m) \over r} -  w \right) - \pi^*(w)$ is increasing with respect to $w$ on $(0, m)$.}

\pf  This difference is increasing if and only if its derivative is positive, which is equivalent to
$$
1 - {\psi'(w) \, \psi'''(w) \over (\psi''(w))^2} > B_1 - 1.  \eqno(4.33)
$$
Rewrite this inequality by setting $w = \hat \psi'(y)$ to obtain
$$
y \, \hat \psi'''(y) < (B_1 - 2) \, \hat \psi''(y).  \eqno(4.34)
$$
After substituting for $\hat \psi$ and simplifying, we observe that the inequality in (4.34) is equivalent to
$$
(B_2 - 2) \left( {y \over y_m} \right)^{B_2 - 2} < (B_1 - 2) \left( {y \over y_m} \right)^{B_2 - 2},  \eqno(4.35)
$$
which is true because $B_2 < B_1$.  \qed

The increasing difference between the investment strategies helps explain the increasing difference between the probabilities of ruin.  For the problem considered by Young (2004), consumption remained constant regardless of the maximum wealth.  Therefore, the agent in that case does not force wealth to stay below the current maximum $m < c(m)/r$, and the corresponding investment strategy entails investing a positive amount of wealth in the risky asset as wealth approaches $m$.  On the other hand, for the paper considered in this paper, consumption remains constant {\it only} because the agent does not allow wealth to increase above the current maximum.  If the agent were to allow wealth to increase above this level, then the permanent increase in the rate of consumption would be too large in the sense that the probability of ruin would increase.  Therefore, the agent in this case forces wealth to stay below the current maximum by investing nothing in the risky asset as wealth approaches $m$.  Propositions 4.6 and 4.7 tell us that this positive difference in the investment strategies is greatest as $w$ reaches $m$ and is smallest for wealth close to $0$, that is, far from $m$.

\subsect{4.3  The Case for which Increasing Maximum Wealth is Optimal}

When wealth $w$ reaches the current maximum wealth $m_0$, the agent either allows wealth to increase above this level or does not.  In Sections 4.1 and 4.2, we studied the case for which increasing the maximum wealth and thereby ratcheting consumption is {\it not} optimal.  We showed that when $m_0 < c(m_0)/r$, then the agent will not allow his wealth to increase above $m_0$ if and only if the condition in (4.23) holds.  Therefore, if it does not hold, then the agent will increase his current maximum, if possible.  We summarize this in the next theorem.

\th{4.8} {If $m_0 < c(m_0)/r$ and if $c(m_0) - m_0 \, c'(m_0) > - \la/\Psi_w(0, m_0),$ then it is optimal for the agent to allow wealth to increase above $m_0$ to $m^*,$ in which
$$
m^* = \inf \{m > m_0: m \ge c(m)/r \hbox{ or } c(m) - m \, c'(m) \le - \la/\Psi_w(0, m) \}.  \eqno(4.38)
$$
Here, $\Psi$ is as in Section $4.1,$ and we explicitly denote that it depends on the variable $m$.}

It follows that the minimum probability of ruin at $(w, m)$ does {\it not} equal $\Psi$ for $m_0 \le m < m^*$; in fact, we have $\psi(w, m) < \Psi(w)$, and we anticipate that Proposition 4.4 and Corollary 4.5 hold in this case.  The classical solution of the following boundary-value problem equals the minimum probability of ruin $\psi$, according to Theorem 2.1.  On $0 \le w \le m$ and $m_0 \le m \le m^*$, $\psi$ uniquely solves
$$
\left\{ \eqalign{& \la f = (r w -c(m)) f_w + \min_\pi \left[ (\mu - r) \pi f_w + {1 \over 2} \sigma^2 \pi^2 f_{ww} \right], \cr
&f(0, m) = 1,  \quad f_m(m, m) = 0, \cr
&\lim_{w \to m^*} f_w(w, m^*)/f_{ww}(w, m^*) = 0.} \right.  \eqno(4.37)
$$
The last condition ensures that $\pi^*$ approaches zero as wealth approaches $m^*$.

The following is an informal discussion of how the interested reader might solve the problem in (4.37) with $m^*$ given by (4.38).   Because the ODE in (4.37) is fully non-linear, consider the dual formulation by hypothesizing that $\psi$ is convex with respect to $w$ and by defining its concave dual $\tilde \psi$ via the Legendre transform.
$$
\tilde \psi(y, m) = \min_w \Big( \psi(w, m) + wy \Big).  \eqno(4.38)
$$
Note that we can recover $\psi$ from $\tilde \psi$ via
$$
\psi(w, m) = \max_y \left( \tilde \psi(y, m) - wy \right).  \eqno(4.39)
$$
One can show that $\tilde \psi$ solves the same ODE as in (4.3) for $(y, m) \in (y_m(m), y_0(m)) \times (m_0, m^*)$, with the following free-boundary conditions for $y_m(m)$ and $y_0(m)$:
$$
\tilde \psi_m(y_m(m), m) = 0, \quad \tilde \psi_y(y(m), m) = m, \quad y_m(m^*) \, \tilde \psi_{yy}(y_m(m^*), m^*) = 0, \eqno(4.40)
$$
and
$$
\tilde \psi(y_0(m), m)) = 1, \quad \tilde \psi_y(y_0(m), m)) = 0.  \eqno(4.41)
$$

As in Section 4.1, $\tilde \psi$ is given by
$$
\tilde \psi(y, m) = \tilde D_1(m) \, y^{B_1} + \tilde D_2(m) \, y^{B_2} + {c(m) \over r} \, y, \eqno(4.42)
$$
in which $B_1$ and $B_2$ are given in (4.5) and (4.6), respectively, and $\tilde D_1(m)$ and $\tilde D_2(m)$ are functions of $m$ to be determined.  The boundary conditions imply that
$$
\tilde D'_1(m) \, y_m(m)^{B_1} + \tilde D'_2(m) \, y_m(m)^{B_2} + {c'(m) \over r} \, y_m(m) = 0,  \eqno(4.43)
$$
$$
\tilde D_1(m) \, B_1 \, y_m(m)^{B_1 - 1} + \tilde D_2(m) \, B_2 \, y_m(m)^{B_2 - 1} + {c(m) \over r} = m,  \eqno(4.44)
$$
$$
\tilde D_1(m) \, y_0(m)^{B_1} + \tilde D_2(m) \, y_0(m)^{B_2} + {c(m) \over r} \, y_0(m) = 1,  \eqno(4.45)
$$
and
$$
\tilde D_1(m) \, B_1 \, y_0(m)^{B_1 - 1} + \tilde D_2(m) \, B_2 \, y_0(m)^{B_2 - 1} + {c(m) \over r} = 0.  \eqno(4.46)
$$

Solve equations (4.45) and (4.46) for $\tilde D_1(m)$ and $\tilde D_2(m)$ to get
$$
\tilde D_1(m) = - {B_2 \over B_1 - B_2} {1 \over y_0(m)^{B_1}} - {c(m) \over r} {1 - B_2 \over B_1 - B_2} {1 \over y_0(m)^{B_1 - 1}},  \eqno(4.47)
$$
and
$$
\tilde D_2(m) =  {B_1 \over B_1 - B_2} {1 \over y_0(m)^{B_2}} - {c(m) \over r} {B_1 - 1 \over B_1 - B_2} {1 \over y_0(m)^{B_2 - 1}}.  \eqno(4.48)
$$
Substitute these expressions into equation (4.44) to obtain
$$
\eqalign{ {c(m) - rm \over r} &= {1 \over y_0(m)} \, {B_1 B_2 \over B_1 - B_2} \left[ \left( {y_m(m) \over y_0(m)} \right)^{B_1 - 1} - \left( {y_m(m) \over y_0(m)} \right)^{B_2 - 1} \right]  \cr
& \quad + {c(m) \over r} \left[ {B_1 (1 - B_2) \over B_1 - B_2} \left( {y_m(m) \over y_0(m)} \right)^{B_1 - 1} + {(B_1 - 1) B_2 \over B_1 - B_2} \left( {y_m(m) \over y_0(m)} \right)^{B_2 - 1} \right].}  \eqno(4.49)
$$
Differentiate (4.47) and (4.48) with respect to $m$ and substitute the results into equation (4.43) to get
$$
\eqalign{ & {y'_0(m) \over y_0(m)} \left[ \left( {y_m(m) \over y_0(m)} \right)^{B_1 - 1} - \left( {y_m(m) \over y_0(m)} \right)^{B_2 - 1} \right] \left\{ {B_1 B_2 \over B_1 - B_2} \, {1 \over y_0(m)} + {c(m) \over r} \, {(B_1 - 1)(1 - B_2) \over B_1 - B_2} \right\} \cr
& \quad = {c'(m) \over r} \left\{ {1 - B_2 \over B_1 - B_2} \left( {y_m(m) \over y_0(m)} \right)^{B_1 - 1} + {B_1 - 1 \over B_1 - B_2} \left( {y_m(m) \over y_0(m)} \right)^{B_2 - 1} - 1 \right\}. }  \eqno(4.50)
$$
We have $m^*$ from (4.38), and note that $y_0(m^*)$ and $y_m(m^*)$ are given in Section 4.1 with $m = m^*$.  One can numerically solve (4.49) and (4.50) for $y_0(m)$ on $[m_0, m^*]$.  Then, given $y_0(m)$, one can get $\tilde D_1(m)$ and $\tilde D_2(m)$ from (4.47) and (4.48), respectively.  Finally, that gives $\tilde \psi(y, m)$ from (4.42) and $\psi(w, m)$ from (4.39).

\medskip

\centerline{\bf Acknowledgments} \medskip

The first author thanks the National Science Foundation for financial support under grant number DMS-0604491.  The second author thanks the Cecil J. and Ethel M. Nesbitt Professorship for financial support.

\sect{References}


\noindent \hangindent 20 pt   Bayraktar, E. and V. R. Young (2007), Correspondence between lifetime minimum wealth and utility of consumption, {\it Finance and Stochastics}, 11 (2): 213-236.

\smallskip \noindent \hangindent 20 pt  Bayraktar, E. and V. R. Young, ``Minimizing probability of ruin and a game of stopping and control,'' working paper, Department of Mathematics, University of Michigan.



\smallskip \noindent \hangindent 20 pt Dybvig, P. H. (1995), Duesenberry's ratcheting of consumption: optimal dynamic consumption and investment given intolerance for any decline in standard of living, {\it Review of Economic Studies}, 62 (2): 287-313.




\smallskip \noindent \hangindent 20 pt \O ksendal, B. and A. Sulem (2004), {\it  Applied Stochastic Control of Jump Diffusions}, Springer, New York.


\smallskip \noindent \hangindent 20 pt Young, V. R. (2004), Optimal investment strategy to minimize the probability of lifetime ruin, {\it North American Actuarial Journal}, 8 (4): 105-126.



\bye